\documentclass{egpubl}
\usepackage{DH2025}%
 
 \ConferenceSubmission   
\WsPaper           

\usepackage[T1]{fontenc}
\usepackage{dfadobe}  
\usepackage{placeins}
\usepackage{booktabs}
\usepackage{subcaption}

\usepackage{cite}  
\BibtexOrBiblatex
\electronicVersion
\PrintedOrElectronic
\ifpdf \usepackage[pdftex]{graphicx} \pdfcompresslevel=9
\else \usepackage[dvips]{graphicx} \fi

\usepackage{egweblnk} 

\title[Comparing OCR Pipelines for Folkloristic Text Digitization]%
      {Comparing OCR Pipelines for Folkloristic Text Digitization}

\author[O.M. Machidon \& A.L. Machidon]
{\parbox{\textwidth}{\centering O.\,M. Machidon$^{1}$\orcid{0000-0003-3133-1008}
        and A.L. Machidon$^{1}$\orcid{0000-0002-9330-3865} 
        }
        \\
{\parbox{\textwidth}{\centering $^1$University of Ljubljana, Faculty of Computer and Information Science, Slovenia
       }
}
}

\begin{document}


\maketitle
\begin{abstract}
   The digitization of historical folkloristic materials presents unique challenges due to diverse text layouts, varying print and handwriting styles, and linguistic variations. This study explores different optical character recognition (OCR) approaches for Slovene folkloristic and historical text digitization, integrating both traditional methods and large language models (LLMs) to improve text transcription accuracy while maintaining linguistic and structural integrity. We compare single-stage OCR techniques with multi-stage pipelines that incorporate machine learning-driven post-processing for text normalization and layout reconstruction. While LLM-enhanced methods show promise in refining recognition outputs and improving readability, they also introduce challenges related to unintended modifications, particularly in the preservation of dialectal expressions and historical structures. Our findings provide insights into selecting optimal digitization strategies for large-scale folklore archives and outline recommendations for developing robust OCR pipelines that balance automation with the need for textual authenticity in digital humanities research. 


\end{abstract}  

\section{Introduction}
\label{sec:introduction}

The digitization of historical folkloristic texts is a key priority in digital heritage research, enabling long-term preservation, accessibility, and computational analysis of cultural narratives. However, these materials present unique challenges for automated transcription due to their diverse physical and linguistic properties. Folkloristic collections often include a mixture of typewritten and handwritten documents, irregular layouts, archaic or dialectal language, and visual artifacts such as marginal notes or embedded illustrations~\cite{fleischhacker2025enhancing}. These factors complicate the use of standard optical character recognition (OCR) pipelines, which are typically trained on contemporary printed text with relatively clean layouts.

The accuracy of OCR is critical in the digital humanities, where text fidelity directly impacts linguistic analysis, pattern detection, and structural interpretation. In folkloristics, even small errors—such as the misrecognition of dialectal variants or disrupted narrative structures—can lead to misinterpretation of cultural meanings. Consequently, researchers require OCR solutions that balance high throughput with sensitivity to textual authenticity.

Recent advances in OCR and language technologies have introduced new possibilities for improving digitization outcomes. Traditional OCR engines, such as Tesseract or Kraken, have been adapted for historical documents with varying degrees of success, often requiring layout-specific preprocessing and custom model training. In parallel, the emergence of large language models (LLMs) has opened up opportunities for post-OCR correction, normalization, and formatting~\cite{kanerva2025ocr}. LLMs such as ChatGPT~\footnote{https://chatgpt.com/} and Claude~\footnote{https://chat.chatbot.app/} can refine noisy transcriptions and reconstruct coherent paragraphs, but they also raise concerns about textual hallucinations and unintended modernization, which may obscure the original character of folkloric texts.

This paper presents a comparative evaluation of OCR pipelines for the digitization of Slovenian folkloristic materials, with a focus on two main approaches: a single-stage pipeline using the open-source tool olmOCR, and a two-stage pipeline combining Tesseract with LLM-based post-processing. Through a series of experiments on typewritten fairy tales, historical newspapers, and children's magazines, we assess the impact of scan quality, document layout, and language variation on OCR accuracy and usability. Our goal is to provide practical insights and recommendations for digital heritage practitioners seeking scalable, accurate, and context-sensitive solutions for folklore digitization.

The contributions of this study are threefold:
\begin{enumerate}
    \item  a comparison of OCR pipelines on authentic folkloristic corpora in Slovene,
    \item an evaluation of LLM-assisted post-processing with respect to both readability and linguistic authenticity, and
    \item guidelines for selecting or customizing OCR workflows based on document characteristics and scholarly requirements. 
\end{enumerate}
This work contributes to ongoing efforts in the digital humanities to reconcile automation with cultural preservation, advancing best practices for large-scale, historically informed text digitization.
\section{Related Work}
\label{sec:related}

\subsection{OCR for Historical and Folkloristic Texts}

Digitizing historical and folkloristic texts via OCR is a well-studied yet challenging task. Such materials often exhibit \textit{diverse layouts}, including multi-column pages, marginalia, and irregular text blocks, which pose difficulties for automatic layout analysis~\cite{fleischhacker2025enhancing}. Degraded print quality (faded ink, stains, bleed-through) and unusual typography (e.g., 19th-century fonts like Fraktur) further contribute to high OCR error rates~\cite{kanerva2025ocr}. Handwritten folkloric manuscripts add another layer of complexity due to handwriting variation, requiring specialized handwriting recognition models. Additionally, these texts frequently contain dialectal or archaic language not seen in modern training data, confounding standard OCR engines that assume contemporary vocabulary and spelling. As a result, off-the-shelf OCR methods on folkloristic archives can produce noisy output, and fully automatic recognition remains an open challenge in digital heritage preservation~\cite{kanerva2025ocr,fleischhacker2025enhancing}.

Over the past decade, numerous OCR systems and pipelines have been developed to address these challenges. Tesseract---an open-source engine---and the Transkribus platform are widely used for printed and handwritten historical documents, respectively, and are often considered baseline solutions in the cultural heritage community~\cite{fleischhacker2025enhancing}. These general-purpose tools offer pretrained models for many languages; however, they struggle with the peculiarities of historical folkloristic material, such as complex page layouts or antiquated fonts, often requiring significant manual intervention to achieve acceptable accuracy~\cite{fleischhacker2025enhancing}. Even with modern OCR technology, pages with decorative borders or interleaved song lyrics and commentary may need human-guided segmentation to avoid recognition errors~\cite{fleischhacker2025enhancing}.

More specialized OCR approaches have emerged. The OCR4all toolkit integrates advanced layout analysis and model training into an end-to-end workflow, achieving top-tier accuracy on recent historical OCR benchmarks~\cite{fleischhacker2025enhancing}. Likewise, the open-source Kraken OCR engine, designed with historical and non-Latin scripts in mind, has demonstrated superior accuracy on challenging texts compared to Tesseract in many cases~\cite{martinek2020building}. Kraken’s trainable neural OCR models have proven effective on early print typefaces and manuscripts, especially with domain-specific training data. For example, ensemble LSTM OCR models like Calamari have reduced the character error rate on 19th-century Fraktur prints below 1\%, outperforming both Tesseract and commercial engines~\cite{reul2018state}.

Several large-scale heritage projects have adopted customized OCR pipelines to handle domain-specific challenges. For printed folklore compilations, it is common to fine-tune OCR models on sample pages from a collection. For handwritten archives, platforms like Transkribus enable tailored handwritten text recognition (HTR) models for each scribe or region, improving accuracy over generic models~\cite{fleischhacker2025enhancing}. Recent literature shows a clear trend toward domain-adaptive OCR in the humanities: researchers combine layout-specific processing, specialized engines, and post-correction methods to handle the varied nature of historical folklore texts. Our work situates itself in this context by comparing such OCR pipeline components and configurations on folkloristic material.

\subsection{Language Model-Assisted Post-Processing and Normalization}

Given that raw OCR outputs can contain substantial errors, post-processing is often applied to improve text quality. Earlier approaches used dictionaries, spelling variation databases, or statistical models to correct obvious errors. In recent years, however, neural language models have become the preferred method for automatic error correction~\cite{kanerva2025ocr}. For example, the best-performing systems in the ICDAR 2019 competition on post-OCR correction~\footnote{https://icdar2019.org/competitions-2/} used a two-step process: a Transformer-based model (BERT) detected probable errors, followed by a sequence-to-sequence model for corrections~\cite{kanerva2025ocr}.

Large Language Models (LLMs) have been explored as a more powerful alternative~\cite{kanerva2025ocr}. Bourne et al.~\cite{bourne2024clocr} demonstrated that GPT-based correction could reduce character error rates by over 60\% on historical datasets. Recent work has treated historical-to-modern normalization as a translation task, fine-tuning large encoder–decoder models, such as the ``Transnormer,'' to map 18th–19th century texts to contemporary German~\cite{ehrmanntraut2024historical}. These techniques achieved state-of-the-art performance, making texts more accessible for search and analysis.

Despite these advances, researchers have cautioned against potential drawbacks~\cite{kanerva2025ocr,boros2024post}. LLMs, being generative models, may introduce plausible but historically inaccurate content, or over-correct dialectal or archaic expressions. Boros et al.~\cite{boros2024post} found that prompting GPT-4 for OCR correction sometimes decreased overall accuracy due to hallucinated outputs. Even with strong normalization models, perfect automatic standardization remains elusive, with residual error rates reported around 1\%~\cite{ehrmanntraut2024historical}. 

Recent work emphasizes careful integration of LLMs, such as constraining outputs or employing human-in-the-loop validation, to ensure authenticity is preserved~\cite{kanerva2025ocr,boros2024post}.

In summary, the field has evolved from generic OCR engines toward specialized, context-aware OCR pipelines and is now exploring how large language models can refine recognition results. Our work addresses the remaining gap by systematically evaluating complete OCR pipelines on folkloristic corpora, building on prior advancements in both OCR and LLM-assisted post-processing.

\section{Materials and methods}
\label{sec:materials}

\subsection{Data Sources}


To evaluate OCR pipelines on real-world folkloristic and cultural heritage materials, we selected three representative Slovene-language datasets. These sources span different time periods, media formats, and linguistic registers, providing a robust foundation for assessing recognition accuracy, layout sensitivity, and post-processing strategies.

\subsubsection{\textit{Kmetijske in rokodelske novice}}

\begin{figure}[htb]
  \centering
  \includegraphics[width=\columnwidth]{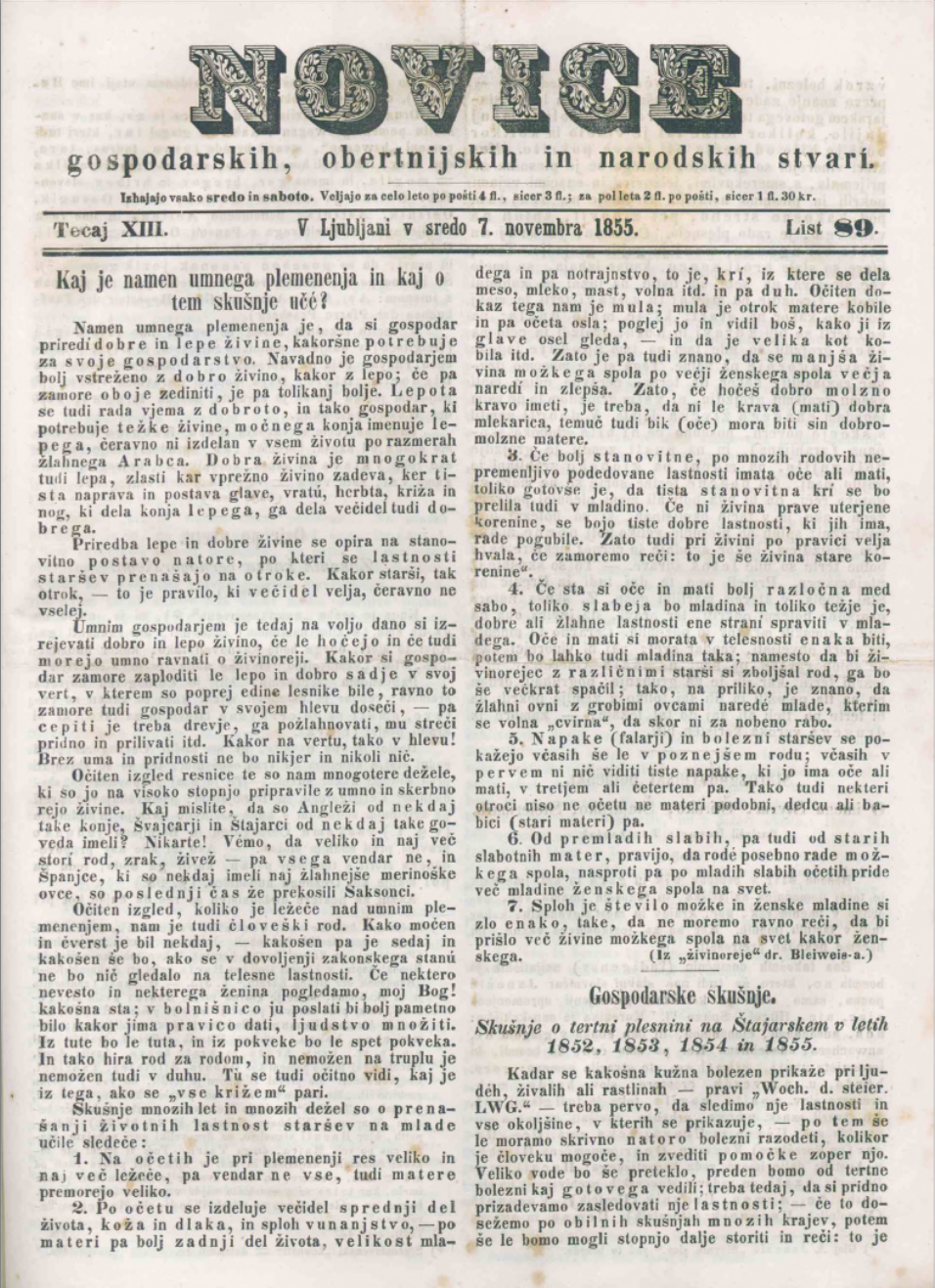}
  \caption{\label{fig:kmetijske_sample}
    A sample page from \textit{Kmetijske in rokodelske novice}, dated November 7, 1855. 
    This historical periodical, written in 19\textsuperscript{th}-century Slovene and printed using antiquated fonts and multi-column layouts, 
    poses significant challenges for OCR due to typographic complexity, degraded print quality, and early orthographic conventions.}
\end{figure}

The first dataset is drawn from the historical periodical \textit{Kmetijske in rokodelske novice}~\cite{kmetijske_novice}, a prominent 19\textsuperscript{th}-century Slovene-language newspaper. Published between 1843 and 1902, this source played a central role in standardizing the Slovene language and promoting literacy and civic education among rural populations. As shown in Figure~\ref{fig:kmetijske_sample}, the periodical uses antiquated fonts (including Fraktur and early Antiqua), multi-column layouts, and historical orthography. These characteristics present significant OCR challenges, including segmentation errors, character confusion, and reduced accuracy on linguistic variants. This dataset is well-suited for evaluating OCR robustness on early printed heritage texts.


\subsubsection{\textit{Ciciban}}

\begin{figure}[htb]
  \centering
  \includegraphics[width=\columnwidth]{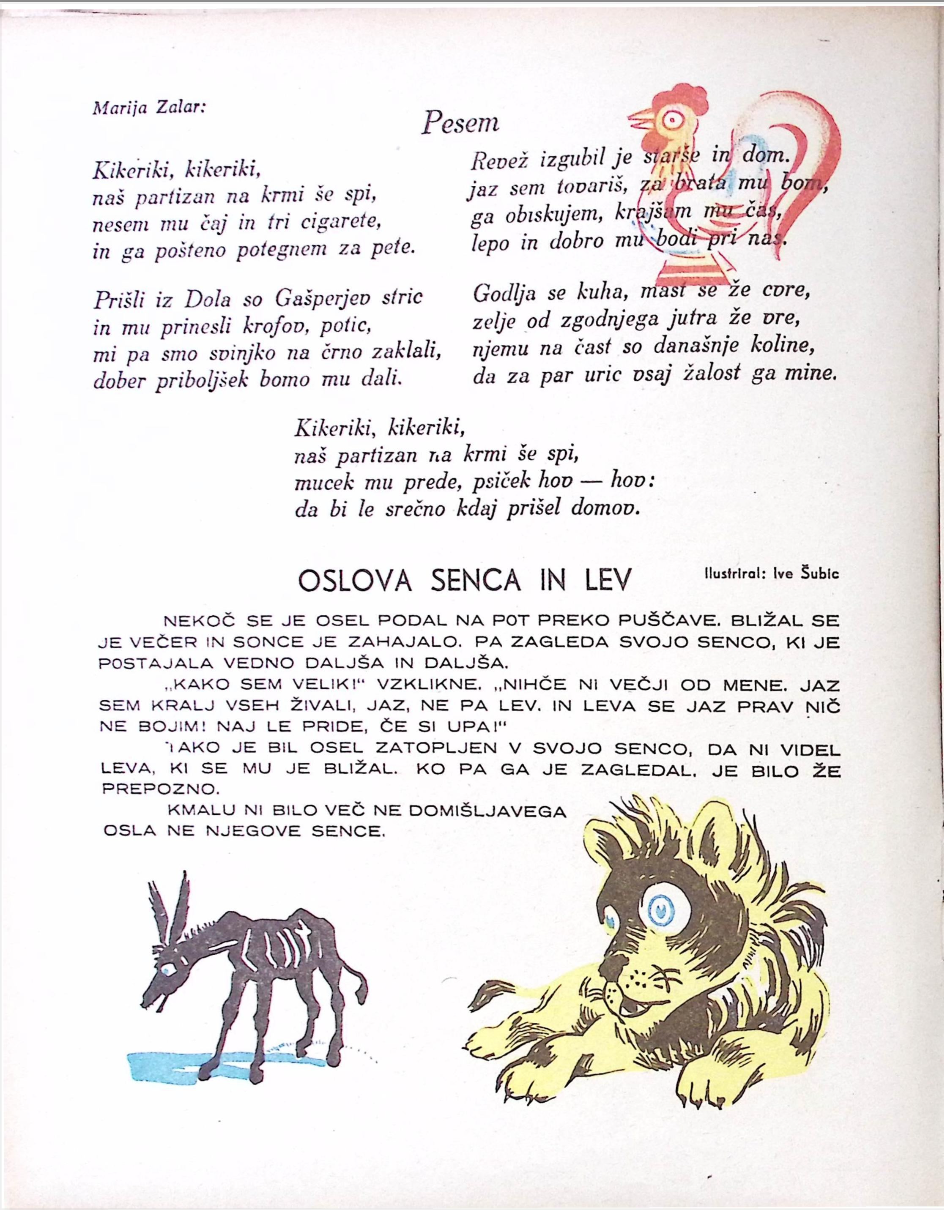}
  \caption{\label{fig:ciciban_sample}
    A sample page from the children's magazine \textit{Ciciban}, featuring a combination of poetic text, short narrative, and illustrations. 
    Documents like this pose layout-specific OCR challenges due to mixed content (e.g., stylized fonts, embedded images, decorative headings), 
    as well as language aimed at young readers. The visual richness of the page exemplifies the need for robust layout analysis 
    and content-sensitive post-processing in OCR workflows.}
\end{figure}

The second dataset comprises issues from \textit{Ciciban}~\cite{ciciban_collection}, a children's magazine first published in 1945 and still in circulation today. Targeted at early readers, the magazine blends poems, short stories, games, and rich visual content. Figure~\ref{fig:ciciban_sample} presents an example page combining illustrated verse and narrative prose. OCR on this corpus is complicated by the interplay of text and images, decorative and variable fonts, and non-linear layout elements. It is especially valuable for testing layout-aware OCR systems and assessing how post-processing models handle stylistically diverse texts intended for pedagogical or literary purposes.

\subsubsection{Inštitut za narodopisje – skenirane pravljice}

\begin{figure}[htb]
  \centering
  \includegraphics[width=\columnwidth]{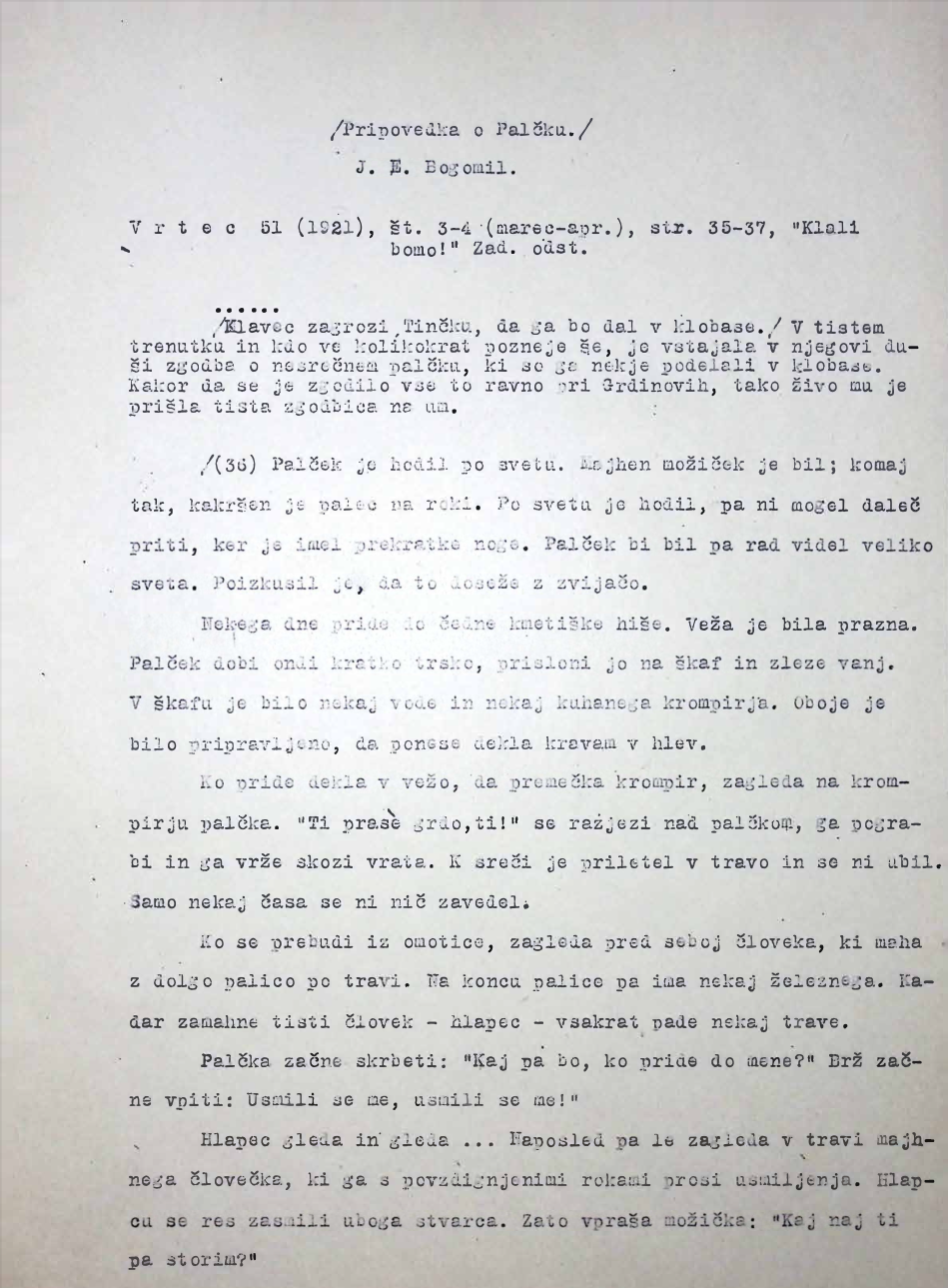}
  \caption{\label{fig:typewritten_sample}
    Sample page from the \textit{Inštitut za narodopisje – skenirane pravljice} collection. 
    The document is typewritten and shows typical layout and character spacing of mid-20\textsuperscript{th} century folkloristic transcriptions. 
    Despite visual uniformity, OCR can be challenged by faded ink, uneven typewriter impressions, and page-level degradation.}
\end{figure}

The third corpus used in this study consists of typewritten folkloristic stories collected by the \textit{Inštitut za narodopisje} (Institute of Folklore Studies) of the Slovene Academy of Sciences and Arts. These materials, created in the mid-20\textsuperscript{th} century, document oral narratives transcribed using typewriters and later digitized as scanned images. A sample page is shown in Figure~\ref{fig:typewritten_sample}. 

This collection presents several interesting challenges for OCR development. Although the documents are typewritten, they vary in quality due to degradation of the paper, ink fading, and scanning inconsistencies. Additionally, the typewriter fonts themselves often introduce visual noise (e.g., irregular spacing or smudged glyphs), which complicates character recognition. The texts sometimes contain handwritten annotations, underlines, or editorial marks, further affecting the recognition accuracy.


\subsection{OCR Pipeline and Processing Workflow}
We evaluated two distinct OCR pipelines to assess their suitability for digitizing Slovene-language folkloristic and historical materials: a single-stage pipeline based on olmOCR and a multi-stage workflow combining Tesseract OCR with large language model (LLM)-based post-processing.

\paragraph*{Pipeline A: olmOCR (Single-Stage)}

The first pipeline employs olmOCR, an open-source toolkit developed by the Allen Institute for AI for high-throughput conversion of PDFs and scanned documents into plain text~\cite{poznanski2025olmocr}. olmOCR~\footnote{https://olmocr.allenai.org/blog} leverages a fine-tuned 7B vision-language model and a novel document-anchoring prompting technique to extract text in natural reading order while preserving structured content such as sections, tables, equations, and handwritten text. It is optimized for scalable, GPU-based batch processing, with an estimated cost of under \$190 per million pages, making it suitable for large-scale digitization workflows.

olmOCR was trained primarily on English academic and technical documents and is not explicitly adapted for Slovene. Nonetheless, in our experiments, it produced surprisingly robust results on typewritten Slovene folkloristic texts and relatively clean layouts. A key advantage is that it operates entirely locally, avoiding third-party API calls and maintaining full data privacy. However, it exhibited limitations when handling documents with complex visual structure (e.g., multi-column formats or image-overlaid text), occasionally misordering content or omitting layout-sensitive elements. Additionally, while olmOCR includes heuristics to reduce hallucinations, some semantic drift was observed on degraded inputs.

\paragraph*{Pipeline B: Tesseract + LLM (Multi-Stage)}

The second pipeline adopts a modular, multi-stage approach. In the first stage, scanned pages are processed using the Tesseract OCR engine with Slovene language support. The resulting raw OCR text is then passed to a large language model (LLM) for post-processing. We used OpenAI’s ChatGPT-4 (GPT-4.o version model, accessed through the OpenAI API~\footnote{https://openai.com/api/}) to correct OCR-induced errors (e.g., fragmented words, spacing inconsistencies, misrecognized characters) and to reconstruct coherent paragraph structure. Prompt engineering was applied to minimize semantic drift and hallucinations. While this post-processing step improved readability and structural clarity, it occasionally led to unintended normalization of dialectal or archaic terms, raising concerns about fidelity in cultural heritage contexts.

To improve performance on documents with complex or visually rich layouts—such as Ciciban—we optionally tested the integration of LayoutParser prior to OCR. This tool was used to detect and segment visual elements (e.g., text blocks, images, headings), enabling a more logical reading order for downstream processing. However, its impact on overall accuracy was limited in practice and did not consistently resolve all layout-related issues.

\paragraph*{Image Preprocessing}

Additional preprocessing experiments were conducted on selected samples using standard enhancement techniques such as grayscale conversion, binarization, and dilation~\footnote{Dilation is a morphological image processing operation that expands or thickens the shapes of objects in a binary image. In OCR preprocessing, it is sometimes used to close small gaps in characters or connect fragmented parts of text to improve recognition accuracy.}
. However, for degraded documents like Kmetijske in rokodelske novice, these steps did not significantly improve OCR accuracy. In some cases, preprocessing introduced artifacts that worsened recognition quality. As a result, most evaluations were conducted using raw scans or lightly optimized images.

\paragraph*{Output Evaluation}

For all pipelines, the output was manually reviewed for recognition accuracy, layout preservation, and language integrity. OCR performance was compared across document types to assess robustness in the presence of historical typography, irregular structure, and linguistic variation. Where relevant, we evaluated the impact of LLMs not only on surface-level error correction, but also on the preservation of culturally meaningful features such as dialectal vocabulary, narrative flow, and typographic cues.

\section{Results and discussion}
\label{sec:results}

\begin{figure}[htb]
  \centering
  \begin{subfigure}{0.48\columnwidth}
    \includegraphics[width=\linewidth]{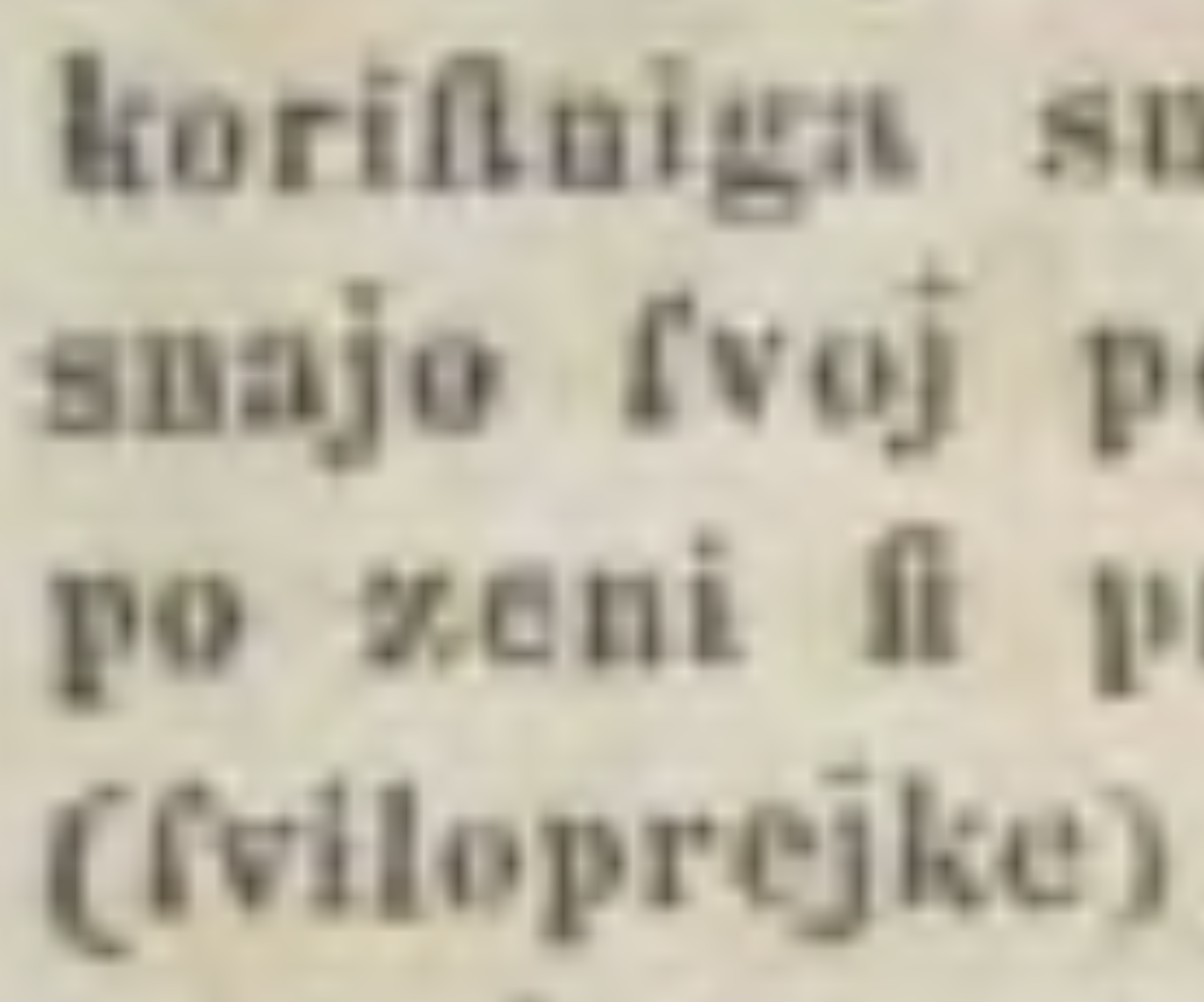}
    \caption{Poor scan quality}
    \label{fig:poor_scan}
  \end{subfigure}
  \hfill
  \begin{subfigure}{0.48\columnwidth}
    \includegraphics[width=\linewidth]{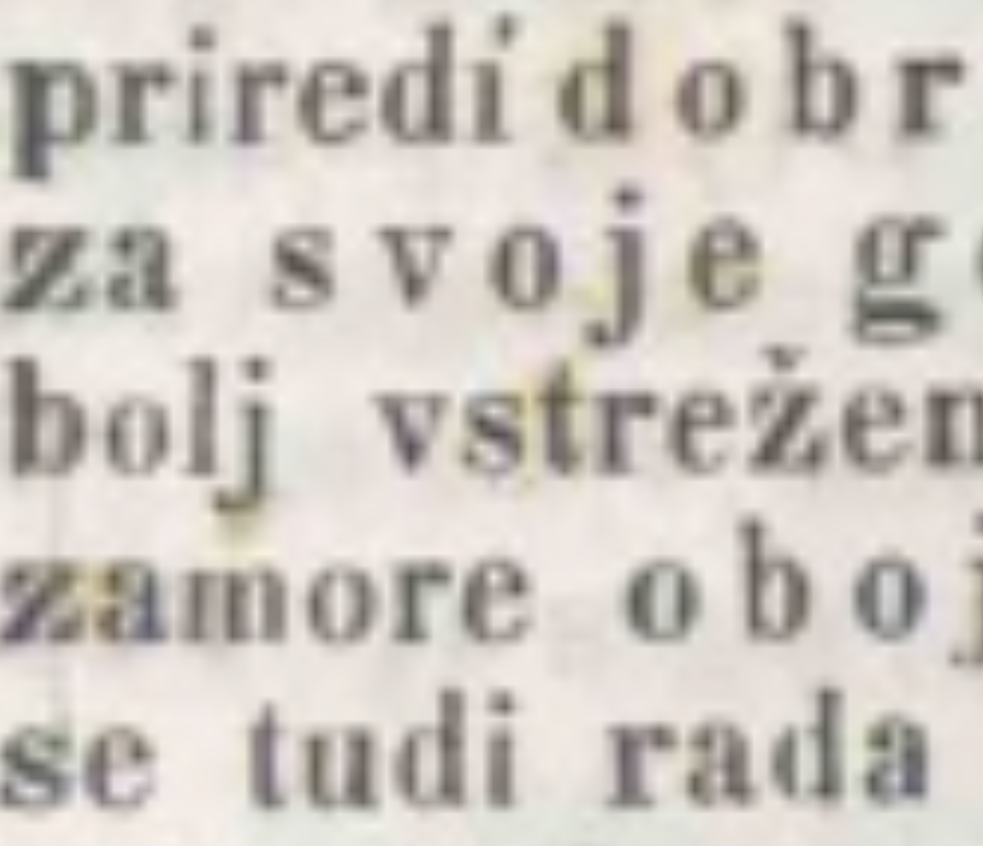}
    \caption{Acceptable scan quality}
    \label{fig:acceptable_scan}
  \end{subfigure}
  \caption{Comparison of scan quality. Left: low-quality scan with noise and degradation; Right: higher-quality scan enabling more accurate OCR results.}
  \label{fig:scan_quality_comparison}
\end{figure}

OCR performance was significantly influenced by the quality of the scanned documents. As shown in the experiments on the Kmetijske in Rokodelske novice dataset, documents scanned at low quality produced OCR results that were largely unusable, while higher-quality scans enabled acceptable transcription accuracy even with minimal preprocessing (for an example of different scan qualities on this dataset see Figure~\ref{fig:scan_quality_comparison}). Attempts to improve lower-quality scans using traditional image preprocessing techniques (grayscale conversion, binarization, dilation) did not yield better results, and in some cases even worsened recognition. This suggests that advanced preprocessing techniques, such as adaptive histogram equalization or deep learning-based enhancement models, may be needed for robust OCR of low-quality historical scans.

\subsection{Comparison of OCR Approaches}

\subsubsection{olmOCR (Single-Stage Pipeline)}

The single-stage pipeline based on olmOCR offered significant advantages in terms of simplicity and privacy. Since olmOCR can be deployed locally, documents were processed without involving third-party services. OlmOCR demonstrated good robustness across different materials: for instance, in the case of the typewritten fairy tales, it produced cleaner and more usable outputs compared to Tesseract-based pipelines. It was particularly effective on relatively simple layouts. However, olmOCR showed limitations when processing documents with complex structures, such as the Ciciban magazine, occasionally misordering text blocks due to its limited layout analysis capabilities.

\subsubsection{Tesseract + LLM (Two-Stage Pipeline)}

The two-stage approach employed Tesseract as the OCR engine, followed by a post-processing stage using ChatGPT-4 or other LLMs (such as Gemini and Claude) for error correction and formatting. This method enabled corrections of common OCR issues, such as fragmented words, missing punctuation, and inconsistent spacing. Furthermore, integration with LayoutParser provided a mechanism for analyzing and reconstructing document layouts, partially mitigating problems with complex page structures.

However, this pipeline also introduced new challenges. LLM-based post-processing occasionally altered the text content beyond correcting errors, replacing dialectal or archaic words with their modern equivalents (e.g., storjice became zgodbe). This "normalization" effect, while improving readability, posed a risk of losing linguistic authenticity, which is critical for folklore studies. Additionally, minor hallucinations were observed, especially when text fragments were highly degraded or incomplete.

\subsection{Evaluation on Different Document Types}

\subsubsection{Folkloristic Newspapers (\textit{Kmetijske in rokodelske novice})}
\begin{figure}[!htb]
  \centering
  \includegraphics[width=\columnwidth]{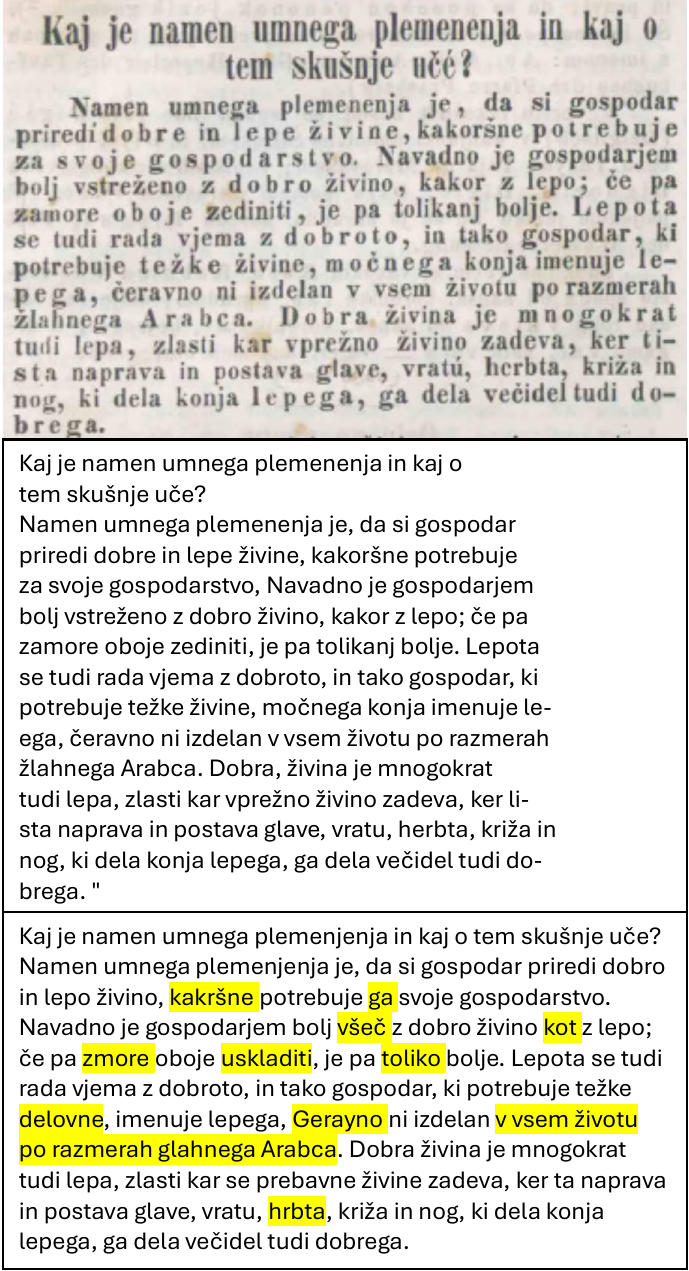}
  \caption{\label{fig:kmet_gpt_comparison}
    OCR and post-processing comparison on a historical page from \textit{Kmetijske in rokodelske novice}. 
    Top: original scan. Middle: raw OCR output using Tesseract. Bottom: ChatGPT-4 enhanced output. 
    Yellow highlights indicate tokens where ChatGPT modified the original OCR text—some represent helpful corrections 
    (e.g., spacing, grammar), while others reveal unintended substitutions or normalizations that may impact authenticity.}
\end{figure}

On degraded historical newspaper scans, both pipelines faced difficulties. olmOCR showed moderate performance, but Tesseract+LLM achieved better readability after post-correction. However, as shown in Figure~\ref{fig:kmet_gpt_comparison}, LLMs sometimes replaced historical terms with modern or unrelated words (e.g., \textit{žlahnega} $\rightarrow$ \textit{glavnega}), which may introduce semantic drift. More specialized OCR training or validation-aware post-processing would be necessary for reliable digitization of these materials.

\subsubsection{Children’s Magazines (\textit{Ciciban})}

\begin{figure}[htb]
  \centering
  \includegraphics[width=\columnwidth]{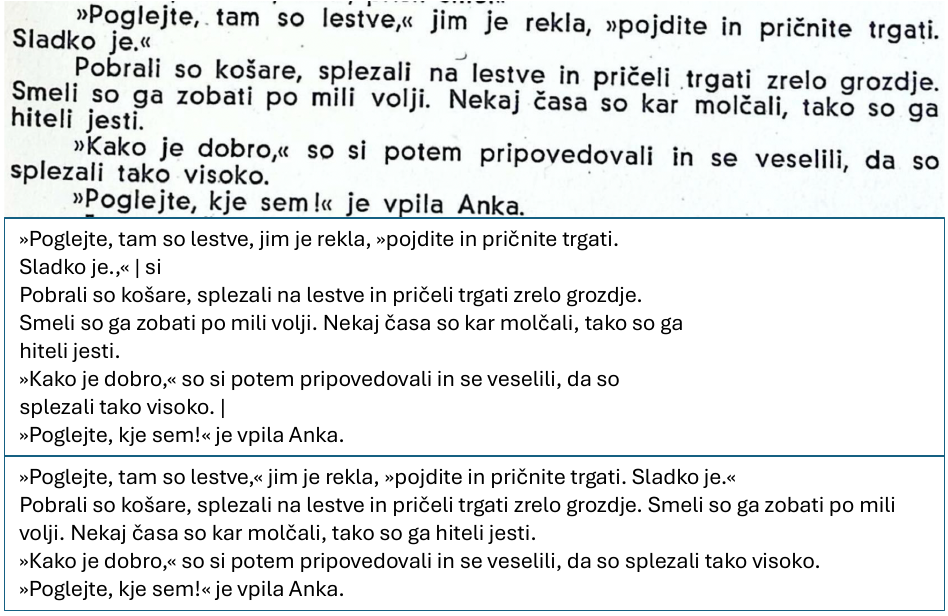}
  \caption{\label{fig:ciciban_gpt_comparison}
    Comparison between raw Tesseract OCR and ChatGPT-enhanced output for a passage from \textit{Ciciban}. 
    The scan quality and font are clean and modern, allowing Tesseract to transcribe the text with high accuracy. 
    ChatGPT-4 further improves readability by restoring correct sentence structure, resolving broken lines, and fixing punctuation—without introducing hallucinations or modifying the original meaning.}
\end{figure}

The Ciciban corpus, consisting of mid- and late-20th-century children's periodicals, features high-quality scans and modern, typewritten fonts. These characteristics allowed Tesseract to achieve strong OCR performance with minimal character-level errors. When combined with ChatGPT-4 in a two-stage pipeline, the output quality improved further: the LLM corrected formatting issues such as broken lines and inconsistent punctuation, while maintaining the semantic integrity of the text. As shown in Figure~\ref{fig:ciciban_gpt_comparison}, the post-processed output was more readable and coherent, without any hallucinations or modernizing alterations.

While olmOCR also produced reasonable results on Ciciban, it lacked sufficient layout awareness and occasionally misordered paragraphs or merged unrelated text segments—particularly in visually complex issues that featured poetry, prose, and images in non-linear arrangements. In such cases, incorporating LayoutParser into the Tesseract-based pipeline helped recover reading order and segment content more effectively. This makes the Tesseract+LLM pipeline, optionally augmented with LayoutParser, the most reliable solution for processing Ciciban documents with mixed layout structures.

\subsubsection{Typewritten Fairy Tales}

\begin{figure}[htb]
  \centering
  \includegraphics[width=\columnwidth]{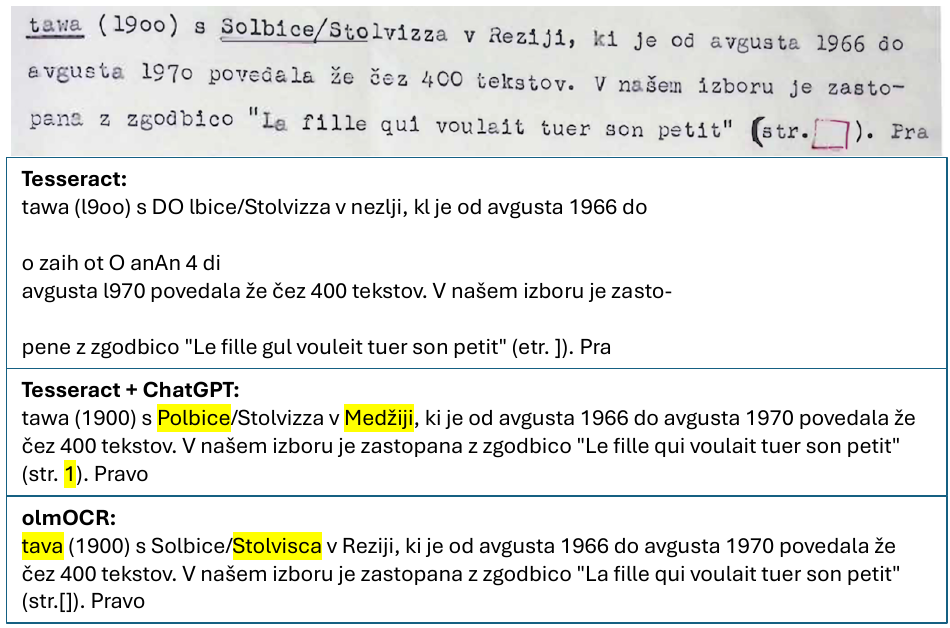}
  \caption{\label{fig:typewritten_comparison}
    Comparative OCR output for a sample from the typewritten fairy tale corpus. 
    Tesseract produced a low-quality raw transcription with significant errors in names, diacritics, and word structure. 
    ChatGPT-4 post-processing attempted to restore coherence but introduced additional hallucinations. 
    In contrast, olmOCR preserved the correct linguistic form with fewer distortions and greater fidelity to the original.}
\end{figure}

For the corpus of mid-20\textsuperscript{th}-century typewritten fairy tales, olmOCR consistently outperformed the Tesseract+LLM pipeline. Tesseract produced noisy raw OCR with numerous errors in character recognition, punctuation, and names—particularly affecting diacritics and borrowed terms. When passed to ChatGPT-4 for post-processing, these issues were only partially corrected. In some cases, the LLM introduced new hallucinations or fabricated plausible-sounding alternatives, which diverged from the original meaning. In contrast, olmOCR yielded more accurate transcriptions directly, with lower hallucination rates and better preservation of the original sentence structure. A representative comparison is shown in Figure~\ref{fig:typewritten_comparison}.

\subsection{Summary of Findings}

\begin{table}[htb]
\centering
\caption{Summary of OCR pipeline performance by dataset}
\label{tab:summary_comparison}
\renewcommand{\arraystretch}{1.05}
\begin{tabular}{p{7.4cm}}
\toprule
\textbf{Kmetijske in rokodelske novice} (19th century, historical newspapers, degraded scans) \\
Preferred pipeline: Tesseract + LLM \\
Strengths: Recovers degraded text, improves readability \\
Weaknesses: Some semantic drift, normalization of historical terms \\
Notes: Minor hallucinations; simple layouts help \\
\midrule
\textbf{Ciciban} (mid-20\textsuperscript{th} century, children's magazine, complex layout) \\
Preferred pipeline: Tesseract + LLM (+LayoutParser) \\
Strengths: High accuracy, effective formatting fixes \\
Weaknesses: Layout inversion risk without parser \\
Notes: Low hallucination risk; layout parsing recommended \\
\midrule
\textbf{Typewritten fairy tales} (mid-20\textsuperscript{th} century, typewritten stories, clean scans) \\
Preferred pipeline: olmOCR \\
Strengths: High fidelity, few hallucinations \\
Weaknesses: Sensitive to scan quality \\
Notes: Best for simple, uniform text \\
\bottomrule
\end{tabular}
\end{table}

Our evaluation across the three distinct folkloristic datasets demonstrates that OCR pipeline performance varies considerably depending on document layout complexity, scan quality, and language characteristics. A summary of our findings is presented in Table~\ref{tab:summary_comparison}. Specifically:

\begin{itemize}
    \item \textbf{Tesseract+LLM} outperformed other approaches on well-scanned historical newspapers (Kmetijske), where ChatGPT-4 helped reconstruct readable text from initially noisy OCR outputs. However, this came at the cost of occasional hallucinations and normalization of historical terms.
    
    \item For Ciciban, which features high scan quality and typographic clarity but complex layouts, the Tesseract+LLM pipeline, optionally enhanced with LayoutParser, provided the most reliable results. Tesseract handled the base recognition effectively, and ChatGPT-4 improved formatting without semantic distortion.
    
    \item On typewritten fairy tales, olmOCR outperformed the Tesseract-based pipeline. Tesseract produced poor raw transcriptions, which the LLM could not reliably correct—sometimes compounding errors with hallucinations. In contrast, olmOCR generated more accurate results.
    
    \item Scan quality remained a critical factor across all datasets. No pipeline was able to reliably salvage degraded or poorly digitized material through conventional preprocessing alone.
\end{itemize}

We conclude that no single pipeline is universally optimal. Instead, a document-sensitive strategy is recommended: in our case, the use olmOCR for clean, typewritten texts, deploying Tesseract+LLM for materials with higher layout complexity and language variation, and applying LayoutParser selectively when complex visual structure threatens reading order.

\section{Conclusions}
\label{sec:conclusions}

This study compared OCR pipeline configurations for the digitization of historical and folkloristic Slovene texts, focusing on three distinct datasets: historical newspapers (\textit{Kmetijske in rokodelske novice}), children’s magazines (\textit{Ciciban}), and typewritten fairy tales. Our experiments evaluated both a single-stage pipeline using \textbf{olmOCR} and a multi-stage pipeline combining \textbf{Tesseract OCR}, \textbf{ChatGPT-4 post-processing}, and optionally \textbf{LayoutParser} for layout analysis.

Our findings show that pipeline performance depends heavily on document characteristics. For high-quality, typewritten materials, olmOCR proved more effective, offering good transcription fidelity with minimal hallucinations. In contrast, Tesseract struggled with initial recognition, and LLM-based correction sometimes amplified errors. For historical newspapers, the Tesseract+LLM pipeline was more successful in reconstructing coherent output, though it introduced some normalization and semantic drift. For richly formatted children’s magazines, Tesseract+LLM again performed well—especially when augmented with layout-aware preprocessing to resolve reading order and structure.

While LLMs significantly enhance formatting and readability, they must be applied with caution to avoid distorting historically and linguistically meaningful content. Similarly, scan quality remains a critical determinant of OCR success; no post-processing approach could fully recover degraded input.

We conclude that no single OCR pipeline is universally optimal. Instead, best practices in folklore digitization call for a tailored strategy. Future work should prioritize the development of domain-adapted OCR models and the integration of validation mechanisms that ensure semantic authenticity during post-correction with LLMs.

\section*{Acknowledgments}
The research presented in this paper was funded by the Slovenian Research Agency through the Gravitacija project ``Veliki jezikovni modeli za digitalno humanistiko'' (GC-0002).

\bibliographystyle{eg-alpha-doi} 
\bibliography{bibligraphy}    

\newcommand{\etalchar}[1]{$^{#1}$}
\begin{thebibliography}{\uppercase{RSWP18}}

\bibitem[BER{\etalchar{*}}24]{boros2024post}
\textsc{Boros E., Ehrmann M., Romanello M., Najem-Meyer S., Kaplan F.}:
\newblock Post-correction of historical text transcripts with large language models: An exploratory study.
\newblock In \emph{Proceedings of the 8th joint SIGHUM workshop on computational linguistics for cultural heritage, social sciences, humanities and literature (LaTeCH-CLfL 2024)} (2024), Association for Computational Linguistics, pp.~133--159.

\bibitem[Bou24]{bourne2024clocr}
\textsc{Bourne J.}:
\newblock {CLOCR-C}: Context leveraging {OCR} correction with pre-trained language models.
\newblock \emph{arXiv preprint arXiv:2408.17428} (2024).

\bibitem[Cic45]{ciciban_collection}
\textsc{Ciciban}:
\newblock Ciciban magazine collection.
\newblock \url{https://www.dlib.si/details/URN:NBN:SI:DOC-1R8Y8Z7L}, from 1945.
\newblock Digitized by the National and University Library of Slovenia (NUK).

\bibitem[Ehr24]{ehrmanntraut2024historical}
\textsc{Ehrmanntraut A.}:
\newblock Historical {G}erman text normalization using type-and token-based language modeling.
\newblock \emph{arXiv preprint arXiv:2409.02841} (2024).

\bibitem[FKG25]{fleischhacker2025enhancing}
\textsc{Fleischhacker D., Kern R., G{\"o}derle W.}:
\newblock Enhancing {OCR} in historical documents with complex layouts through machine learning.
\newblock \emph{International Journal on Digital Libraries 26}, 1 (2025), 3.

\bibitem[KLKG25]{kanerva2025ocr}
\textsc{Kanerva J., Ledins C., K{\"a}pyaho S., Ginter F.}:
\newblock {OCR} error post-correction with llms in historical documents: No free lunches.
\newblock \emph{arXiv preprint arXiv:2502.01205} (2025).

\bibitem[kme02]{kmetijske_novice}
Kmetijske in rokodelske novice.
\newblock \url{https://nl.ijs.si/imp/nuk/dl/NUKP14041-1843.html}, (1843–1902).
\newblock Digitized by the National and University Library of Slovenia (NUK).

\bibitem[MLK20]{martinek2020building}
\textsc{Mart{\'\i}nek J., Lenc L., Kr{\'a}l P.}:
\newblock Building an efficient {OCR} system for historical documents with little training data.
\newblock \emph{Neural Computing and Applications 32} (2020), 17209--17227.

\bibitem[PBD{\etalchar{*}}25]{poznanski2025olmocr}
\textsc{Poznanski J., Borchardt J., Dunkelberger J., Huff R., Lin D., Rangapur A., Wilhelm C., Lo K., Soldaini L.}:
\newblock olm{OCR}: Unlocking trillions of tokens in pdfs with vision language models.
\newblock \emph{arXiv preprint arXiv:2502.18443} (2025).

\bibitem[RSWP18]{reul2018state}
\textsc{Reul C., Springmann U., Wick C., Puppe F.}:
\newblock State of the art optical character recognition of 19th century fraktur scripts using open source engines.
\newblock \emph{arXiv preprint arXiv:1810.03436} (2018).

\end{thebibliography}


\end{document}